\documentstyle[12pt,epsf]{article}

\def\hybrid{
      \topmargin-20pt      
      \oddsidemargin 0pt
      \headheight 0pt   \headsep 0pt
     \textwidth 6.5in  % US paper
     \textheight 9in         % US paper
      \textwidth 6.25in % A4 paper
      \textheight 9.5in % A4 paper
      \marginparwidth .875in
      \parskip 5pt plus 1pt   \jot = 1.5ex}
\hybrid

\def\x{\times}
\def\ox{\otimes}
\def\o+{\oplus}
\def\ra{\rightarrow}
\def\lra{\longrightarrow}
\def\da{\downarrow}
\def\beqa{\begin{eqnarray}}
\def\eeqa{\end{eqnarray}}

\newcommand{\un}{\underline}

\newcommand{\ti}{\tilde}
\newcommand{\al}{\alpha}
\newcommand{\de}{\delta}
\newcommand{\ep}{\epsilon}
\newcommand{\si}{\sigma}
\newcommand{\th}{\theta}
\newcommand{\om}{\omega}

\newcommand{\A}{{\cal A}}
\newcommand{\B}{{\cal B}}
\newcommand{\E}{{\cal E}}
\newcommand{\G}{{\cal G}}
\newcommand{\cO}{{\cal O}}
\newcommand{\cP}{{\cal P}}
\newcommand{\resetcounter}{\setcounter{equation}{0}}

\begin{document}

\thispagestyle{empty}
\rightline{LMU-TPS 04/15}
\rightline{hep-th/0412182}
\vspace{2truecm}
\centerline{\bf \LARGE Standard Model bundles}
\vspace{0.3truecm}
\centerline{\bf \LARGE of the heterotic string}

\vspace{1.5truecm}
\centerline{Gottfried Curio$^*$\footnote{curio@theorie.physik.uni-muenchen.de}}
\vspace{.6truecm}

\vspace{.4truecm}
\centerline{{\em Department f\"ur Physik, Ludwig-Maximilians-Universit\"at
M\"unchen}}
\centerline{{\em Theresienstra\ss e 37, 80333 M\"unchen, Germany}}

\vspace{1.0truecm}
%%%%%%%%%%%%%%%%%%%%%%%%%%%%%%%%%%%%%%%%%%%%%%%%%%%%%%%%%
%\vspace{.4truecm}
\begin{abstract}
We show how to construct supersymmetric three-generation models 
with gauge group and matter content of the Standard Model 
in the framework of non-simply-connected 
elliptically fibered Calabi-Yau manifolds $Z$. 
The elliptic fibration on a cover Calabi-Yau, where the model 
has $6$ generations of $SU(5)$
and the bundle is given via the spectral cover description,
has a second section 
leading to the needed free involution.
The relevant involution on the defining spectral data of the bundle
is identified for a general Calabi-Yau space of this type 
and invariant bundles are generally constructible.
\end{abstract}

\bigskip \bigskip
\newpage

\section{Introduction}

A search for the gauge group $G_{SM}$
and chiral matter content of the Standard model 
within the framework of string theory 
may be done under different perspectives.
We adopt as guidelines here the following principles: 
the number and respective ranks of the 
gauge group factors should not be put in just by hand
(like collecting different gauge groups just in the 
manner needed) but rather come in one package (as from a broken GUT group); 
gravity should be on the same level as the gauge field theory 
(and not in a different sector
but rather should come united with the gauge forces like in a
string compactification model); with an eye on the stabilization of moduli
(which has to be achieved later when discussing the continuous parameters 
of the Standard model) the discussion should not be restricted from the outset 
to having important geometric moduli fixed (by working in some 
geometric limit of the space) 
but rather should be carried out at a generic point in moduli space
so that it retains some flexibility 
when including further requirements; finally the construction 
should be as simple as possible in the given framework
(so that especially it remains manageable when proceeding further).

Efforts to get a (supersymmetric) phenomenological spectrum from 
the $E_8\times E_8$ heterotic string on a Calabi-Yau $Z$
started with embedding the spin connection in the gauge connection 
which gave an unbroken $E_6$ (times a hidden $E_8$
which couples only gravitationally). Then a further breaking
of the gauge group can be achieved 
by turning on Wilson lines using a non-trivial
$\pi_1 (Z)$. Though the simplest constructions of Calabi-Yau spaces 
have a trivial fundamental group
one can still produce a non-trivial $\pi_1$ by dividing 
$Z$ by a freely acting group $G= \pi_1$, provided such an operation 
exists. This leads at the same time to a reduction by $|G|$
of the often large number of generations $\chi(Z)/2$.

This approach was generalised [\ref{W}] to the case of embedding 
instead of the tangent bundle an $SU(n)$ bundle for $n=4$ or $5$,
leading to unbroken $SO(10)$ resp. $SU(5)$ of even greater 
phenomenological interest than $E_6$. This subject was revived when
the bundle construction was made much more explicit for the case of
elliptically fibered Calabi-Yau $\pi: Z\ra B$ [\ref{FMW}]. In subsequent work 
this extended ansatz showed among other things to have a much 
greater flexibility in providing one with three-generation models 
of the corresponding unbroken GUT group [\ref{C}]. 

There remains to go to the Standard Model gauge group in this 
framework. 
For the Hirzebruch surfaces $F_m$, $m=0,1,2$ as bases $B$
the special elliptic fibrations $\pi: Z\ra B$ we will have to 
consider are smooth. 
However none of them has non-trivial $\pi_1(Z)$
(the Enriques surface as base $B$ has $\pi_1(B)={\bf Z}_2$ leading to 
a non-trivial $\pi_1(Z)$ but, as pointed out by E. Witten, 
does not lead to a three generation model, cf. [\ref{DLOW}]). 

The elliptic framework can nevertheless 
give a three generation model of Standard Model 
gauge group and matter content by working with an $SU(5)$ bundle
leading
to a $SU(5)$ gauge group on a space admitting a free involution
$\iota$ which after modding it out gives 
a smooth Calabi-Yau $Z'=Z/{\bf Z_2}$
\beqa
Z & \stackrel {\rho}{\lra}  & Z'\nonumber\\
\pi \da & & \; \da \, \pi'\\
B & \lra & B'\nonumber
\eeqa
On $Z'$ one can turn on a ${\bf Z_2}$ Wilson line 
with generator ${\bf 1_3} \o+ {\bf - 1_2}$
breaking $SU(5)$ to $G_{SM}$
\beqa
SU(5) \lra SU(3) \, \x \, SU(2) \, \x \, U(1)
\eeqa

On the other hand to actually compute the generation number one
has to work 'upstairs' as $Z'$ does not have a section but only a bi-section 
(left over from the two sections of $Z$) and so one can not 
use the spectral cover method there directly\footnote{The bundle
is described fibrewise (where it decomposes into a sum of line bundles)
by a collection of points, for which a choice of 
reference point $p$ (the "zero" in the group law) has to be made (to represent 
a degree zero line bundle by the divisor $Q_i-p$); 
to have the reference point globally one needs a section.}.
\beqa
V & \lra & V'\nonumber\\
\da & & \da \\
Z & \stackrel {\rho}{\lra} & Z'\nonumber
\eeqa
If $V'$ over $Z'$ is the searched for 3-generation bundle
then the bundle $V=\rho^*V'$ on $Z$ has $6$ generations
and is 'moddable' by construction.
Conversely, having constructed a bundle above on $Z$ with $6$ 
generations, one assures that it can be modded out by $\iota$ 
(to get the searched for bundle on $Z'$) by demanding that $V$ 
should be $\iota$-invariant. 
So one needs to specify on $Z$ an $\iota$-invariant 
$SU(5)$ bundle that leads to $6$ generations
and fulfills some further requirements of the spectral cover construction,
essentially a bound on the $\eta$ class expressing the five-brane 
effectivity required from anomaly cancellation.

As it will be our goal to 'mod' not just the Calabi-Yau spaces but 
also the geometric data describing the bundle (and this transformation
of bundle data into geometric data uses in an essential way the 
elliptic fibration structure) we will search only for actions which
preserve the fibration structure, i.e. $\underline{\iota}\cdot \pi=\pi\cdot 
\iota$ 
with $\underline{\iota}$ an action on the base.  
\beqa
Z & \stackrel {\iota}{\lra}  & Z\nonumber\\
\pi \da & & \; \da \, \pi\\
B & \stackrel {\underline{\iota}}{\lra} & B\nonumber
\eeqa
This has the consequence that our elliptically fibered Calabi-Yau spaces 
will actually have $2$ sections\footnote{We will use the same notation 
for a section, its image and its cohomology class.} 
$\sigma_1$ and $\sigma_2=\iota \si_1$
($B$-model spaces).  

Turning this around we will look for an elliptically fibered
Calabi-Yau $Z$ with a changed type of elliptic fibre so that the
global fibration has then besides the usually assumed single section 
($A$-model) a second one ($B$-model);
this will then lead to a free involution $\iota$ on $Z$.
The elliptic Calabi-Yau spaces over Hirzebruch surfaces 
we will actually consider are close cousins
of the "$STU$"-models of Hodge numbers $(3,243)$ prominent in the
$N=2$ string duality (there are versions of this space over 
the Hirzebruch surface ${\bf F_m}$ for $m=0,1,2$ with the same Hodge numbers). 
For example over ${\bf F_2}$ the exchange of the
$A$-fibre ${\bf P}_{2,3,1}(6)$ with the $B$-fibre ${\bf P}_{1,2,1}(4)$ 
leads one from the ${\bf P}_{1,1,2,8,12}(24)$ to the 
${\bf P}_{1,1,2,4,8}(16)$ of $h^{1,1}=4$ and $h^{2,1}=148$
(one has again three versions of this space for $m=0,1,2$; we stick 
to these $m$ as the only ones for which the global $B$-model is smooth). 
Demanding the existence of an appropriate involution 
(fibrewise the translation in the group law
with $\si_2$, i.e. $\si_2$ has to be at a $1/2$-division point)
leaves only $75$ deformations.

In section $2$ the $B$-model spaces along with their cohomological 
data are introduced; for $F_0$, $F_2$ we are writing down a free involution. 
Then the spectral cover construction
of bundles is recalled and the Chern classes of $V$ and
the $\iota$-action for our Calabi-Yau are described. 
Requiring $c_1(V)=0$ fixes $V$ in terms of an $\eta \in H^{1,1}(B)$ 
and a {\em pair} $(\mu,\lambda)$ of half-integers occuring in the 
twisting bundle $L$ on the spectral cover $C=n\si_1+\eta$
\beqa
\label{L type}
c_1(L) &=& 
\Big( 
(n\si_1 + \eta + c_1 ) /2 + \lambda \, ( n\si_1 - \eta + nc_1) + \mu \, \ep 
\Big) \; |_C
\nonumber\\
&=&\Big( x\, \si_1 + \al + \mu \, \ep \Big) \; |_C
\eeqa
(with $\al \in H^2(B)$).
Here the new invariant $\mu$ occurs as one can use now 
the new class $\delta= \mu \, \ep |_C= \mu (\si_1 - \si_2 + c_1)|_C$
in the $B$-model,
to get a broader class of bundles for which we identify the $\iota$-action; 
then $c_2(V)$ is given with its $\mu$-dependence
\beqa
c_2(V)=(1-\mu)\eta \si_1 + \mu \eta \si_2 - \mu (1-\mu) \eta c_1 + kF
\eeqa
(with classes of the form $aF$ 
which contain the parameter $\mu$ kept seperate) and 
examples of a $6$ generation model ('above') are given
(the effectivity bound on $\eta$ is discussed in appendix C). 
Finally the action of $i$ on $c_2(V)$ is read off 
which amounts just to the parameter substitution 
$\phi_{\mu}: \mu \ra 1 - \mu$; 
this suggests already (what is formally demonstrated in appendix A
and constitutes the main result of this paper) 
that $\phi_{\mu}$ gives the action $I$ of the involution 
also on the bundle itself and not just on the second 
Chern-class of $V$
(the relevant bundle $L$ on $C$ comes from restriction to $C$ 
of a line bundle $l=\cO(D)$ on $Z$)
\newpage
\beqa
\label{I operation}
L\;\; \; & \stackrel{F}{\lra} & V \nonumber\\
I \da \;\; & & \da \, \iota \;\;\;\;\;\;\;\;\;\;\;\;\;\;  
I \cO\Big(x\, \si_1 + \al + \mu \, \ep \Big) 
= \cO\Big(x\, \si_1 + \al + (1-\mu) \,\ep \Big) \\
I(L) & \stackrel{F}{\lra} & \iota V \nonumber
\eeqa
Here we suppressed that one 
actually starts with $l$ and first builds $j_* j^* l=j_*L$, where
$j: C\hookrightarrow Z$, 
before applying the Fourier-Mukai transform
(actually we will prove (\ref{I operation}) just restricted to $C$, i.e.
only for $L$ what is sufficient for our purposes); 
building $F(j_* L)$ is the spectral cover construction. 
For ${\bf F_m}$ with $m$ even the invariant element 
$\de_{\mu=1/2}=\ep /2|_C$, i.e. $C$ 
has to be tuned to be an integral
divisor class so that the construction is well-defined. 

As the focus of the present paper is on 
how to construct invariant bundles over a
general elliptically fibered Calabi-Yau and not so much 
on specific examples we do not scan exhaustively all possible bases 
nor do we search specifically for stable $\eta$
%but mention rather only an illustrative example 
because the
invariance question, as is appearent from (\ref{I operation}), 
is independent of specific $\eta$-classes.

Note also that as the ${\bf 10}$ and the 
${\bf \bar{5}}$ will come in the same number of families by anomaly 
considerations it is enough to adjust $c_3(V)/2$ to get all 
the Standard model fermions.

Finally some remarks on technical points and relation to [\ref{ACK}] and 
[\ref{DOPW1999}], [\ref{DOPW2000}]. 
Using the $B$-models our discussion will rely on [\ref{ACK}] 
whose main points we will recall. 
From the generation formula found in [\ref{C}] it is immediate 
to write down the $\eta, \lambda$ parameters for the 
6-generations models above on $Z$. 
There remain the issues of the $\eta$-bound 
(i.e. effectivity of the five-brane class) and $\iota$-invariance. 
The mentioned models were only partially given in [\ref{ACK}] 
as the ones over ${\bf F_2}$ violated an upper bound on the $\eta$-class. 
That actually a less restrictive version of the bound can be used 
and that when demanding $\iota$-invariance 
the bounds are completely unproblematical  
is described here in subsect. \ref{uppbou}.  
In appendix A the $\iota$-action is identified already on the bundle level 
(not just for $c_2(V)$). 
When forcing two sections in the $A$-model and 
resolving the singularities occurring by the moduli specialisation 
(leading to a model similar but  
different from [\ref{ACK}]) one has
to consider individual resolution classes 
whose behaviour under the Fourier-Mukai transform can be difficult to treat 
(one alternative then is to consider more involved constructions 
on a special $Z$ given by the fibre product of two elliptic surfaces). 
A uniform treatment of all these classes is
essentially corresponding to $\delta_{\mu = 1/2}$ 
which is actually to be tuned integral
so that here a construction just with $L$-classes coming from the full $Z$ 
is possible. The mentioned features 
lead to a relatively simple construction. 

\section{The Calabi-Yau spaces with free ${\bf Z_2}$ action} 
\label{spaces} 

\subsection{Change of fibre type} 

To have the extra structure, which allows for free involution  
on the elliptically fibered Calabi-Yau we will use a different  
elliptic curve than the fibre usually taken in the 
Weierstrass $A$-model which we recall first. 

{\em $A$-model} 

\noindent 
The mentioned standard description ($A$-model) has the form 
\beqa 
\;\;\;\;\;\;\;\;\;\;\;\mbox{$A$-fibre} \;\;\;\;\;\;\;\;\;\;\;\;\; 
\;\;\;\;\;\;\;\;\;\;\;\; 
y^2+x^3+z^6+sxz^4=0\;\;\;\;\;\;\;\;\;\;\;\; 
\;\;\;\;\;\;\;\;\;\;\;\; \;\;\;\;\;\;\;\;\;\;\;\; 
\eeqa 
We will be interested in a global version of these descriptions over 
a complex surface 
$B$ so that our Calabi-Yau $Z$ can be described by a 
generalized Weierstrass equation in a ${\bf P}^2$ bundle $W$ over $B$ 
(note that the fibre is not ${\bf P}_{2,3,1}(6)$). 
This gives for the $A$-fibre (cf. [\ref{FMW}]) 
\beqa 
\;\;\;\;\;\;\;\;\;\;\;\mbox{global $A$-model} 
\;\;\;\;\;\;\;\;\;\;\;\;\; 
\;\;\;\;\;\;\;\;\;\;\;\; 
zy^2 + x^3 + a_4 xz^2 + b_6 z^3=0\;\;\;\;\;\;\;\;\;\;\;\; 
\;\;\;\;\;\;\;\;\;\;\;\; \;\;\;\;\;\;\;\;\;\;\;\; 
\eeqa 
where the given variables $x,y,z$ and coefficient functions $a,b$ 
are sections of ${\cal L}^i$, 
with $i=2,3,0$ and $i=4,6$, respectively, 
with the line bundle ${\cal L}=K_B^{-1}$ over $B$. 
One has an obvious projection $\pi: Z \rightarrow B$ and a section 
$\si : B\rightarrow Z$ given by $z=0$. 
The Chern classes of $Z$ are given by (cf. [\ref{FMW}]) 
($c_1, c_2, \alpha$ will always denote $c_1(B), c_2(B), \pi^* \alpha$ 
for $\alpha \in H^{2}(B)$) 
\beqa 
c_2(Z)= 12\; \si \; c_1 +c_2+11c_1^2 \;\;\;\;\;\;\;\;\;\;\; 
, \;\;\;\;\;\;\;\;\;\;\; c_3(Z)=-60c_1^2 
\eeqa 
From the weights of $a_4$ and $b_6$ one gets $9^2+13^2-3-3-1=243$
deformations over ${\bf F_0}$.

{\em $B$-model} 

\noindent 
We will be interested in the so-called $B$-fibre ${\bf P}_{1,2,1}(4)$ 
(cf. [\ref{ACK}]) 
\beqa  
\;\;\;\;\;\;\;\;\;\;\;\mbox{$B$-fibre} \;\;\;\;\;\;\;\;\;\;\;\;\; 
\;\;\;\;\;\;\;\;\;\;\;\; 
y^2+x^4+sx^2z^2+z^4=0\;\;\;\;\;\;\;\;\;\;\;\; \;\;\;\;\;\;\;\;\;\;\;\;  
\;\;\;\;\;\;\;\;\;\;\;\; 
\eeqa 

In the case of the global $B$-model the fibre ${\bf P}^2$ is actually 
a weighted ${\bf P}_{1,2,1}$ with $y$ being a section of 
${\cal O}(2)$. Then the variables $x,y,z$ and coefficient functions $a,b,c$ 
are sections of ${\cal L}^i$ with $i=1,2,0$ and $i=2,3,4$, respectively
and one has a well-defined equation
\beqa
\label{Bmodel}
\;\;\;\;\;\;\;\;\;\;\;\;\mbox{global $B$-model}
\;\;\;\;\;\;\;\;\;\;\;
y^2 + x^4 + a_2 x^2z^2 + b_3 xz^3 + c_4 z^4=0\;\;\;\;\;\;\;\;\;\;\;\;
\;\;\;\;\;\;\;\;\;\;\;\;\;\;\;\;\;\;\;\;\;\;\;\;
\eeqa
From the weights of $a_2$, $b_3$ and $c_4$ one gets $5^2+7^2+9^2-3-3-1=148$
deformations over ${\bf F_0}$.
The generic form of the global $B$-model is smooth [\ref{ACK}]
over a Hirzebruch base
${\bf F_m}$ for $m=0,1,2$, 
{\em the cases we restrict ourselves to from now on}.
Now a second, cohomologically inequivalent section $\si_2$ occurs:
considering the equation at the locus $z=0$, 
i.e. $y^2=x^4$ (after $y\ra iy$)
one finds 8 solutions which constitute two equivalence 
classes in ${\bf P}(1,2,1)$: $(x,y,z)=(1,\pm 1,0)$. 
$y=+1$ corresponds to the zero section (in the group law) $\si_1$, 
while the other one has rank 1 in the Mordell-Weil group, 
i.e. generates infinitely many sections. 
For special points in the moduli space 
we can bring it to a 2-torsion point (in the group law) 
leading to the shift-involution, cf. subsect \ref{freeinvo}.        

Let us keep some relations on record: 
one has the adjunction relations ($i=1,2$)
\beqa
\label{adj rel}
\si_i\, (\si_i + c_1 ) = 0
\eeqa
(with $\si_i:=\si_i(B)$) and, as $\si_1$ and $\si_2$ are disjoint,
$\si_1 \si_2 =0$, the important class
\beqa
\label{epsilon}
\ep : \, = \si_1 - \si_2 + c_1
\;\;\;\;\;\;\;\;\;\;\;\;\;\;\;\; (\;\mbox{with}\;\;\;\;
\si_1 \, \ep =0 \;)
\eeqa
turns out to be trivial on $\si_1$.
Finally one finds for the Chern classes
\beqa
\label{c2B}
c_2(Z)=  6\; (\si_1 + \si_2) \; c_1 +c_2+ 5c_1^2 
\;\;\;\;\;\;\;\;\;\;\; , \;\;\;\;\;\;\;\;\;\;\; c_3(Z)=-36c_1^2
\eeqa
(Over ${\bf F_m}$ cf. $h^{2,1}=148$ above
and $h^{1,1}=4$.)
For a relation between the $A$-model, when enforced to have also
a second section, and the $B$-model cf. appendix B.

{\em Examples over Hirzebruch bases}

%Let us take a closer look at the cohomology of our spaces.
Usually, in the $A$-model, one has $k+1$ $H^2(Z)$-cohomology classes from the 
divisors ($k:=h^{1,1}(B)$), namely $\sigma$ and the $\pi^* \alpha$; 
similarly one finds $k+1$ $H^4(Z)$-cohomology classes from the curves
$F$ (the elliptic fibre) and $\sigma  \alpha$ 
(denoting $\sigma_*  \alpha$ resp. equivalently $\si \cdot \pi^* \al$).

In the $B$-model one has $k+2$ divisor cohomology 
classes (including $\si_2$) and $2k+1$ curve cohomology classes
(including $\si_2\alpha$). So all but one of the new
curve classes $\sigma_2 \alpha$ must be dependent on the other classes.
So one has one relation over $B={\bf F_m}$
whose $H^{1,1}$ is generated 
by the class $b$ of the base-${\bf P^1}$ (of $b^2=-m$)
and $f$ of the fibre-${\bf P^1}$.

From Poincare duality and the intersection products
(with $c_1({\bf F_m})=2b+ (2+m)f$)
\[ \begin{array}{c||c|c|c|c}
  & \sigma_1 & \sigma_2 & \pi^* b & \pi^* f \\
\hline\hline
F         & 1   & 1   & 0  & 0 \\
\sigma_1 b& m-2 & 0   & -m & 1 \\
\sigma_1 f& -2  & 0   & 1  & 0 \\
\sigma_2 b& 0   & m-2 & -m & 1 \\
\sigma_2 f& 0   & -2  & 1  & 0
\end{array} \]
one finds, with $X:= \frac{1}{2}(\sigma_2 - \sigma_1)f$ 
as new independent element 
besides the $F$ and the $\sigma_1 \alpha$, 
\beqa 
\label{diff}
(\sigma_2 - \sigma_1) \, \alpha = (\alpha \cdot c_1)\, X 
\eeqa 
(as is checked directly for $\alpha = xb+yf$; here $X$ is integral). 
Note that despite appearance to the contrary the term $\ep /2$ (like $X$) 
shows here integral intersection pairings: for the class 
$c_1 /2$ as well as the class $(\si_1 - \si_2) /2$ have
for ${\bf F_m}$ with $m$ even, 
integral intersection pairings with the displaeyed classes as can be seen 
from inspecting the intersection table. 

$(\si_1, \si_2)$ or $(\si_1, \ep)$ build a basis for 
$H^2(Z)/H^2(B)$ with $D=x\si_1+y\si_2+\al
=x\si_1-\mu\si_2+\al=\ti{x}\si_1+\mu \ep+\ti{\al}$ where  
$\alpha \in H^2(B)$, $\mu = -y$,
$\ti{x}=x-\mu$ and  $\ti{\al}=\al - \mu c_1$.
With 
\beqa
\th:=\frac{1}{2c_1^2}\ep \, c_1
=\frac{1}{2}(\frac{1}{c_1^2}(\si_1-\si_2)c_1+F)=\frac{1}{2}(-X+F)
\eeqa
one has $y=D\cdot \th$,
i.e. taking intersection with the class
$\th$ is defined in 
$H^2(Z)/H^2(B)$ and $\si_k \cdot \th = \de_{k2}$ resp. 
$\ep \cdot \th =1$. (For a 
one-dimensional base one would have to use 
$\th:= \ep /2$.) 

{\em An involution} 

We define an involution  
$I_D: D_{\mu}\ra D_{1-\mu}$ on the space $H^2(Z)$ in the form 
\beqa
\label{I_D operation}
I_D: \;\;\; \ti{x}\si_1+\mu \ep+\ti{\al}
\lra \ti{x}\si_1+(1-\mu) \ep+\ti{\al}
\eeqa
In the other basis this amounts to
\beqa
\label{involution}
I_D: \;\;\; x\si_1 + y\si_2 + \al
&\lra & x\si_1 + y\si_2 + \ep + 2y \ep +\al
\eeqa
Therefore the affine transformation given by $I_D$ decomposes as follows
\beqa
I_D = T_{\ep} \circ R_{\ep} \;\;\; , \;\;\;\;\; 
R_{\ep}: \; D \lra D + (D \cdot \frac{\ep \, c_1}{c_1^2}) \, \ep
\eeqa
where $T_{\ep}$ is the translation by $\ep$ and the linear part is given by
the reflection $R_{\ep}$ in the $\mu = 1/2$ axis 
in the $(\mu, \tilde{x})$-plane 
(note $R_{\ep}(\ep)=-\ep$ so that $R_{\ep}\neq \iota$ 
as $\iota \, \ep = -\ep + 2 c_1$). 

{\em Remark:} Later
it will be important to understand the classes ($k=1,2$)
\beqa
D_k:= R_{\ep} ( \si_k)=\si_k + 2\de_{2k}\ep
\eeqa
Note that the 
class $D_2:= \si_2 + 2 \ep$ does 
not\footnote{showing that $R_{\ep}$ does not come 
from a map on the space (unlike $\iota$ which keeps irreducibility)}
come from a smooth {\em irreducible}
effective divisor or else it would be a section (as $F\cdot \ep =0$)
and $(D_2 + c_1) D_2=6c_1\ep=48(F-X)\neq 0$ would have to vanish, 
cf. (\ref{adj rel}) (for the situation in the (specialised) $A$-model
cf. appendix B). But this product 
vanishes on (e.g. spectral) surfaces (classes)
like $x\si_1+\al$ what will be enough for our purposes
(cf. remark after (\ref{I operation}) 
and the reasoning in appendix A).

\subsection{\label{freeinvo}
Free ${\bf Z}_2$ action on the global $B$-model $Z$}

We recall how a smooth 
%elliptically fibered Calabi-Yau threefolds 
$Z$ with (fibration-compatible) free involution $\iota$ 
leaving  the holomorphic $(3,0)$-form invariant is found [\ref{ACK}],
so that $Z'=Z/\iota$ is a smooth Calabi-Yau.

In the global $B$-model one has a involutive shift symmetry which is
free at least on the generic fibre. The existence of the shift 
symmetry means that the second section is at a {\em globally}
specified order-$2$ point; this 
in turn means specialization $b=0$ of the complex parameters (the spaces
then turn out to be singular; so this has still to be modified)
\beqa
\label{shift}
\mbox{shift-symmetry}\;\;\;\;\;\;\;\;\;\;\;\;  
(x,y,z)\stackrel {sh}{\longrightarrow} (-x,-y,z)
\;\;\;\;\;\;\;\;\;\;\;\; \mbox{for} \;\;\;\;\;\;\;\;\;\;\;\; b=0
\eeqa
As $\iota$ is compatible with the fibration we
have an involution $\underline{\iota}$ on $B$ with
$\underline{\iota} \cdot \pi=\pi\cdot \iota$.
$Z'=Z/\iota$ is again an elliptic fibration (of $B$-type elliptic fibre) 
over a base $B'=B/\underline{\iota}$. 
To get $\iota$ one starts with the elliptic fibration $Z$ with $2$ 
sections and uses the operation (\ref{shift}) together 
with an in general non-free involution $\underline{\iota}$ on $B$.
As $\iota$ identifies the two sections their image 
downstairs in $Z'$ 
(where thereby an independent divisor class in $H^{1,1}$ is lost)
will be an irreducible surface $\sigma$ (still isomorphic 
to $B$) which is only a bisection: from $pr^* \sigma \cdot 
pr^* f=2\;\, \sigma \cdot f$ where $f$ denotes the fibre downstairs, lying 
over a generic point $b'$ in the base $B'$ of $Z'$, one finds 
$\si \cdot f =2$, the left hand side being $4$ as each of the two sections 
$\sigma_i$ over $\sigma$ intersects each of the two fibers in $Z$ 'above' 
(lying over the 
preimages of $b'$ in $B$) twice.

Over ${\bf F_m}$ the involution $\un{\iota}$ will be chosen as 
a combination of the non-trivial involutions on the two
individual ${\bf P^1}$'s, so $\un{\iota}: (z_1,z_2)\ra (-z_1,-z_2)$ 
in local coordinates. The fix-point locus in the base is 
generically disjoint from the discriminant of the elliptic fibration. 
The operation
over the fix-locus in the base must be free in the fibre, i.e. 
a shift by a torsion point. So one has 
to use actually a fibration where such a shift exists globally (even
if only on a sublocus of moduli space). The fibre shift must be combined
with an operation in the base because
at the singular fibers the pure fibre operation alone 
ceases to be free; also using just the fibre shift would restrict one 
to the locus $b=0$ where $Z$ is singular. 

One has [\ref{ACK}] 
a free involution 
for a specialised $Z$ over $F_0$ and $F_2$ 
($\Omega_3^{holo}$, in its explicit coordinate expression, 
is invariant as $\iota$ involves an even number of minus signs) 
\beqa 
\label{iota} 
(z_1, z_2 \, ; x,y,z)\stackrel{\iota}{\longrightarrow} 
(-z_1, -z_2 \, ; -x,-y,z)
\eeqa
which 
does not necessarily restrict one to the $b=0$ locus
(for $b=0$ it is identical to the shift 
provided by the second section, which is then an involution). 
The involution does not exist on the fibre as such,
but can exist, when combined with a base involution, on a subspace 
of the moduli space (not as 'small' as the locus $b=0$)
where the generic member is still smooth. 
From (\ref{Bmodel})
the coefficient functions should transform under $\un{\iota}$
as $a_2^+, b_3^-, c_4^+$, i.e. over ${\bf F_0}$, say, 
only monomials $z_1^pz_2^q$ within $b_{6,6}$ 
with $p+q$ even are forbidden 
(note that even away from $b=0$ the 
coordinate involution still maps the two sections on another);
similarly in $a_{4,4}$ and $c_{8,8}$ $p+q$ odd is forbidden.
Therefore the number of deformations drops 
to $h^{2,1}=(5^2+1)/2+(7^2-1)/2+(9^2+1)/2-1-1-1=75$.
The discriminant 
%$\Delta=9 a^4 c + 81 a b^2 c-72 a^2 c^2+144 c^3
%- \frac{9}{4}a^3 b^2- \frac{243}{16}b^4$ 
of (\ref{Bmodel})
remains generic since enough terms in $a,b,c$ survive, 
so $Z$ is still smooth. The Hodge numbers $(4,148)$ resp. $(3, 75)$
show that the Euler number is indeed reduced by the factor $2$.

%\newpage

\section{The bundles}

\resetcounter

\subsection{\label{speccov}The spectral cover description}

In the spectral cover description of an $SU(n)$ bundle [\ref{FMW}]
one considers the bundle $V$
first over an elliptic fibre $E$ and then pastes together these descriptions
using global data in the base $B$. Over $E$ 
an $SU(n)$ bundle $V$ over $Z$ (assumed to be fibrewise semistable)
decomposes as a direct sum $\oplus_{i=1}^n {\cal L}_{q_i}$
of line bundles ${\cal L}_{q_i}=\cO_E(q_i - p)$
of degree zero ($p$ the zero element); this is 
described as a set of $n$ points $q_i$ which sum to zero. 
Variation over $B$ gives a hypersurface $C \subset Z$
which is a ramified $n$-fold cover of $B$ given as a locus $w=0$ 
with $w$ a section of ${\cal O}(\sigma)^n\otimes {\cal M}$ 
(with a line bundle ${\cal M}$ over $B$ of class 
$\eta \in H^{1,1}(B)$)
\beqa
C=n\si_1 + \eta
\eeqa

The idea is then to trade in the $SU(n)$ bundle $V$ over $Z$,
which is in a sense essentially a datum over $B$, for a line bundle $L$ over
the $n$-fold (ramified) cover $C$ of $B$: one has 
\beqa
\label{FM}
V=p_*(p_C^*L\otimes {\cal P})
\eeqa
with $p:Z\times_B C\ra Z$ and $p_C: Z\times_B C\ra C$ the projections
and ${\cal P}$ the
global version of the Poincare line bundle over $E_I\times E_{II}$ 
(actually one uses a symmetrized version of this), i.e. 
the universal bundle which realizes $E_{II}$ as  
moduli space of degree zero line bundles over $E_I$. 
${\cal P}=\cO(\Delta - \si_{I,1}\x_B Z_{II} - Z_I \x_B \si_{II,1}-c_1)$
(with $c_1$ denoting here $K_B^{-1}$)
becomes trivial on 
$\si_{I,1}\x_B Z_{II} $ and $Z_I \x_B \si_{II,1}$ and furthermore ($k=1,2$)
\beqa
\label{Poincare restrictions}
 \cP|_{Z \x_B \si_k=Z}=\cO (-\de_{k2}\ep )
\eeqa

A second parameter in the description of $V$ is given by a
half-integral number $\lambda$ as
the condition $c_1(V)=\pi_*\Big(c_1(L)+\frac{c_1(C)-c_1}{2}\Big)=0$ gives
(with $\gamma \in ker\, \pi_*:H^{1,1}(C)\ra H^{1,1}(B)$)
\beqa
c_1(L)=-\frac{1}{2}(c_1(C)-c_1)+\gamma = 
\frac{n\si_1 + \eta + c_1}{2}+\gamma
\eeqa
(for other degrees of freedom cf. [\ref{CD}])
where $\gamma$ is, in the $A$-model, 
being given by ($\lambda \in \frac{1}{2}{\bf Z}$)
\beqa
\gamma_A=\lambda(n\si_1-\eta+nc_1)|_C
\eeqa
as the latter is (in the $A$-model) 
the only generically existent class which projects to zero. 
In the $B$-model one has actually a further possibility (see below).

The most natural requirement (and generically the only one) to assure
integrality of
\beqa
c_1(L)=n(\frac{1}{2}+\lambda)\sigma +(\frac{1}{2}-\lambda)\eta+
(\frac{1}{2}+n\lambda)c_1
\eeqa
is\footnote{More exotic possibilities 
include $\lambda=\frac{1}{2n}, \eta\equiv 0\, {\rm mod}\, n$ for $n$ odd or 
$\lambda=\frac{1}{4}, \eta=2c_1\, {\rm mod}\, 4$ for $n=4$ [\ref{Andreas}].} 
$\lambda \in \frac{1}{2}+{\bf Z}$ for $n$ odd resp. 
$\lambda \in {\bf Z}$, $\eta\equiv c_1 \, {\rm mod} \, 2$ for $n$ even
(then $c_2(V)\in H^4(Z,{\bf Z})$).\\

{\em genuine $B$ type bundles}

Up to now the influence of chosing a $B$-model
Calabi-Yau $Z$ had a rather restricted impact. 
Essentially the influence of this alteration was restricted to the change 
in $c_2(Z)$ whose consequences for the upper bound of $\eta$ will be 
discussed below. A much more interesting new freedom arises because
one has more divisors (the second section) and so a new option arises
to build up a line bundle on the spectral cover 
from a class $\delta=\gamma_B$ 
\beqa
\delta:\,= \mu \ep\, |_C=\mu(\sigma_1+c_1-\sigma_2)|_C
\eeqa
(note that $p_*(\delta)=0$; {\em when considered in $Z$} one would have
$\delta = \mu \, \ep \, C 
= \mu \, \ep \, \eta=\mu \, (\eta \cdot c_1) \, (F-X)$ from (\ref{diff})).

One finds for the most general combination $\gamma=\gamma_A+\delta$ 
to be used in building up $L$ that $\mu$ can be 
integral with $\lambda$ restrictions unchanged;
$\mu$ can be half-integral if $\frac{1}{2}(\si_1 - \si_2 + c_1)|_C$
belongs to the integral cohomology of $C$, i.e. if $C$ can be chosen that way
(as $\ep /2$ was not integral already on $Z$, 
i.e. before pulled-back (restricted) to $C$ by $j: C \hookrightarrow Z$).

\subsection{The cohomology of the bundles}

Grothendieck-Riemann-Roch gives the cohomological data of $V$ 
from those of $L$
\beqa
\label{GRR}
ch(V)Td(Z)=p_*\bigg( e^{c_1(p_C^* L \otimes {\cal P})}Td(Z\x_B C)\bigg)
\eeqa
One has
from (\ref{FM}) and with the relations
$p_*(c_1^2({\cal P}))=-2\eta(\si_1+c_1)$, 
$p_*(\gamma_A c_1({\cal P}))=0$, 
$p_*(\delta c_1({\cal P}))=\mu \, \eta \, \ep$,
$p_*(\gamma_A  \delta)=0$ by (\ref{epsilon}), 
$\pi_* (\gamma_A^2)=-\lambda^2 n \eta (\eta - n c_1)$ 
and $\pi_*\delta^2=-2\mu^2\eta c_1$ that 
%\newpage
%($\gamma = \gamma_A + \delta$)
\beqa
\label{c2V}
c_2(V)&=&p_*\Big[ \frac{(n\si_1 + \eta + c_1)^2}{24}
-\frac{c_2(Z\x_B C)-c_2(Z)}{12}
-\frac{1}{2}\bigg(\gamma + c_1({\cal P})\bigg)^2 \Big]\nonumber\\
&=&-\frac{1}{2}p_*\bigg(c_1^2({\cal P})\bigg)
-p_*\bigg(\delta c_1({\cal P})\bigg)
-\frac{1}{2}p_*(\delta^2)\nonumber\\
&&
-\frac{n}{24}\bigg( (n^2-1)c_1^2+3\eta(\eta-nc_1)\bigg) 
-\eta c_1 -\frac{1}{2}p_*(\gamma_A^2)\nonumber\\
&=&\eta\si_1 -\mu \eta \, \ep +\mu^2\eta c_1-\frac{n^3-n}{24}c_1^2
+\frac{1}{2}(\lambda^2-\frac{1}{4})n\eta (\eta-nc_1)\nonumber\\
&=&(1-\mu)\eta \si_1 + \mu \eta \si_2 - \mu (1-\mu) \eta c_1 + kF
\eeqa
where we have kept seperate the classes of the form $aF$
which contain the parameter $\mu$. \\

{\em The generation number}

Note that the generation number $c_3(V)/2$ is 
unchanged\footnote{That $c_3(V)$ is independent of $\mu$ 
is further explained in [\ref{ACK}] 
where also the fact that $c_2(V)$ is for $\mu=0$ the
same as in the $A$-model, in principle obvious but not manifest 
from (\ref{GRR}), is discussed further.}
because a newly occurring term
$-\frac{3}{2}\mu\eta c_1(\sigma_1-\sigma_2)$ in $c_3(V)/2$ does not contribute
(because it vanishes after integration over $Z$ as both 
$\sigma_i$ are sections leading after integration over the fibre to an 
integral over $B$ times $(1-1)$). 
Concerning chiral matter one finds 
as number of net generations [\ref{C}]
\beqa
\label{generation formula}
\frac{1}{2}c_3(V)=\lambda \eta (\eta-nc_1)
\eeqa
Note that in the case of an $SU(5)$ bundle one has because of the decomposition
\beqa
{\bf 248}=({\bf 5},{\bf 10})\oplus ({\bf 10},{\bf \bar{5}})\oplus
({\bf \bar{5}},{\bf \bar{10}})\oplus ({\bf \bar{10}},{\bf 5})\oplus
({\bf 1},{\bf 24})\oplus ({\bf 24}, {\bf 1})
\eeqa 
to consider also the $\Lambda ^2 V={\bf 10}$ 
(unlike in the case of an $SU(3)$ or $SU(4)$ bundle) to get 
the ${\bf \bar{5}}$ part of the fermions ${\bf 10} \oplus {\bf \bar{5}}$; 
but the ${\bf 10}$ and the 
${\bf \bar{5}}$ will come in the same number of families by anomaly 
considerations.

When searching for a 3-generation model 'below', i.e. 
a 6-generation model 'above',
(\ref{generation formula}) restricts one to 
$\lambda=\pm 1/2$ or $\pm 3/2$ 
where one has then to construct on the Calabi-Yau 'above' (before the modding)
an $SU(5)$ model with $\eta(\eta-5c_1)=\pm 12$ or $\pm 4$.

Examples, over $F_2$, of $6$ generations bundles are
immediately found from formula (\ref{generation formula})
\beqa
\eta =14b+22f \;\;\; , \;\;\; \lambda = + 3/2 \;\;\;\;\;\;\;\;\;\; 
; \;\;\;\;\;\;\;\;\;\;
\eta =24b+30f \;\;\; \lambda = - 1/2
\eeqa

As the focus in this paper is on the involution-invariance condition
we have relegated further discussion of the $\eta$-sector to 
appendix C where further conditions on the $\eta$ class are described
(such as effectivity of the five-brane class occurring
in the anomaly cancellation condition).

\subsection{\label{invbundle}Invariance of the bundles under the involution}

As conditions on $\eta$ and $(\mu, \lambda)$
from $\iota^*V=V$ one finds first $\un{\iota}(\eta_1)=\eta_1$;
as in our case $B=F_0$ or $F_2$
the involution induced on $H^{1,1}(B)$ 
is trivial anyway, this does not represent any restriction:
$b$ and $f$ build a basis of $H^{1,1}(B)$ 
and the top $(1,1)$ form of these
two ${\bf P}_1$ are given in local coordinates by, say, $dzd\bar{z}$; but these
classes are invariant under the operations (actually $(-1)$'s) 
on the ${\bf P}_1$.

Actually also the actual spectral cover surface $C$ should and can easily be 
chosen to be invariant 
under the appropriate induced operation $\iota'$.
As $C$ sits actually
in the dual Calabi-Yau where the shift-symmetry (\ref{shift}) is ineffective
one has here to use (cf. [\ref{DOPW2000}])
a $\iota'$ where $sh$ is split off\footnote{For $b_3=0$ this would be 
$\iota': (z_1,z_2;(x,y,z))\ra (-z_1,-z_2;(-x,y,z))$;
note that then $\iota= sh \; \circ \; \iota'$.}.

Finally and most importantly we describe how 
the operation of the involution on $V$
is expressed on the level of $L$. This operation can be discussed separately 
and at the end as it is independent of the $\eta$-sector.
The relation (\ref{c2V}) for the second Chern class
\beqa
c_2(V)=(1-\mu)\eta \si_1 + \mu \eta \si_2 - \mu (1-\mu) \eta c_1 + kF
\eeqa
suggests
that this operation
is given not by any non-trivial map 
on the classes but just by a parameter substitution 
\beqa
\label{mu invol}
V \longrightarrow \iota^* V  \;\;\;\;\;\;\; \Longleftrightarrow
\;\;\;\;\;\;\; (\mu, \lambda) \longrightarrow (1-\mu, \lambda)
\eeqa
which correctly also leaves invariant the $aF$ term as $a$ is invariant
under the substitution in $\mu$. From (\ref{mu invol}) one notices 
that as this operation, i.e.
(\ref{I operation}) resp. (\ref{I_D operation}),
acts only in the $\ep$-sector it keeps the
general form of $c_1(L)$ (which was fixed by $c_1(V)=0$)
in contrast to a possible action $\si_1 \leftrightarrow \si_2$ (in $L$ !), 
i.e. again just $\iota$,
which would have been also compatible with (\ref{c2V}).

Therefore one finds the critical line
$\mu = 1/2$ 
as invariance locus. Note that this not only makes the cohomology class
$c_2(V)$ invariant; rather, this being just a parameter substitution, 
it suggests (\ref{mu invol}), i.e. that the substitution in $\mu$ already
on the level of the ($V$-defining) line bundle $L$ itself
corresponds to the involution on the bundle $V$ 
\beqa
I \cO\Big(x\, \si_1 + \al + \mu \, \ep \Big) 
= \cO\Big(x\, \si_1 + \al + (1-\mu) \,\ep \Big)
\eeqa
(at least when restricted to $C$ 
what is the relevant case in the application to $L$; 
this relation is checked formally in the appendix A). 
Note further that $C$ has to be chosen such that
the term $\frac{1}{2}(\si_1 - \si_2 + c_1)|_C$
belongs to the integral cohomology of $C$.

I thank B. Andreas and A. Klemm for discussion.

\newpage

\appendix

\section{Appendix: the involution on the bundle data}

\resetcounter

\noindent
We identify the action $I$ of the involution directly on the bundle 
data, specifically on the bundle $L$ on $C$. For this we work on 
$Z \x_B Z$ rather than $Z \x_B C$ and use the language of the Fourier-Mukai 
transform $F$ such that $V=p_*(p_C^*L \ox {\cal P})=F ( j_* L)$ where 
$j:C\ra Z$ is the inclusion. As the line bundles 
$\cO(\si_i), \cO(\eta), \cO(c_1)$
(with the notation $\cO(c_1)=\cO(K_B^{-1})$) 
from which $L$ is built exist already 
on $Z$ one has $L=j^* l$ for a line bundle $l$ on $Z$; in the application
$l=\cO(x \si_1 + \al + \mu \ep)$ 
where $c_1(L)=(x \si_1 + \al + \mu \ep)|_C$. 
Let us now assume
that we have already taken $i^{'} C = C$ (cf. sect. \ref{invbundle}). 
Then, taking into account that
$I j_* j^* l = j_* j^* I l$, 
we will identify $I l$ 
(cf. the analogous case in [\ref{DOPW2000}]) 
where $F I l = i F l$ in general 
(cf. (\ref{I operation}))\footnote{Nevertheless we 
actually prove this only on the $L$-level, i.e. restricted to $C$
what is sufficient.};
more precisely we want to show $I \, l_{\mu}=l_{1-\mu}$, 
i.e. that $I$ here operates as
$I_D$, cf. (\ref{involution}), so that $I$
keeps the class of the relevant $l$'s in (\ref{L type}).
As a locally free sheaf $\cO(\al)$ 
coming from the base can be carried through everywhere as a tensor factor 
we just show (\ref{involution}) in the form ($k=1,2$)
\beqa
\label{Involution}
I \, \cO(z\si_k)=\cO(z\si_k + \ep + 2z\de_{k2} \ep)
\eeqa

Now note that one has $F \G = R p_{1*}(p_2^* \G \ox \cP)$ 
with\footnote{$\cO(c_1)=K_Z\ox \pi^* K_B^{-1}=\omega_{Z/B}$ the dualizing sheaf
of relative Serre duality $(R^1\pi_* \G)^*\cong \pi_*(\G^* \ox \omega_{Z/B})$}
$\hat{F}=R p_{1*}(p_2^* \G \ox \cP \ox \cO(c_1))$ as inverse transform
(when also taking into account an exchange of the roles of $Z_1$ and $Z_2$; 
further a (-1)-shift in the grading is involved here). So $\hat{F}=DFD$ 
($D$ is taking the dual sheaf) 
and we can evaluate $I$ in the form $I=DFD \; i\,  F
%=DF \, i \, DF
$.

Before we do the actual calculation 
we define inductively (the isomorphism class of) a 
rank $a$ vector bundle $V_a$ (for $a\ge 1$) 
by the non-split short exact sequence (SES)
\beqa
\label{Va}
0 \lra \cO(ac_1) \lra V_{a+1} \lra V_a \lra 0
\eeqa
starting with $V_1=\cO(\si_1)$
(and then also $V_0=\si_{1*}\si_1^* \cO(\si_1)=\si_{1*} \cO_B(-c_1)=
\cO_{\si_1}(-c_1)$)
and fulfilling $\pi_* V_a^* = 0$ (i.e. $\cO_B$)
and $R^1\pi_* V_a^* = \cO_B(-ac_1)$.
Concerning the uniqueness 
note that from the Leray spectral sequence of the fibration
one has a SES 
\beqa
0 \ra H^1\Big(B, \pi_*V_a^* \ox \cO_B(ac_1)\Big) \ra
H^1\Big( Z, V_a^*\ox \cO_Z(ac_1) \Big) \ra
H^0 \Big( B, (R^1 \pi_* V_a^*)\ox \cO_B(ac_1)\Big) \ra 0\nonumber
 \eeqa
and\footnote{using $H^1(B, \cO_B(ac_1))\cong
H^1(b, \pi_{F*}\cO_B(ac_1))=0$ 
with $\pi_F: {\bf F_m}\ra {\bf P^1_b}$, cf. [\ref{Hartsh}] 
(although $c_1$ is not ample for ${\bf F_2}$ 
so that the Kodaira vanishing theorem does not apply)} 
the group of extensions 
$\mbox{Ext}^1(V_a, \cO(ac_1))=H^1(Z, V_a^* \ox \cO(ac_1)) 
= H^0(B, \cO_B) = {\bf C}$.

Then one finds indeed 
(we focus on positive $\mu$, i.e.
negative $z$; arguments for positive $z$ are analogous)
with the auxiliary relations (\ref{minus a}) and (\ref{Xa}) 
presented below that ($k=1,  2$; $a > 0$; 
for $a=0$ we will find $F^1 \cO = V_0$)
\newpage
\beqa
F  \,  \cO(-a\si_k) &=& V_a    \ox \cO(-\de_{k2}\ep) [ -1 ]\\
iF \,  \cO(-a\si_k) &=& iV_a   \ox \cO(-\de_{k2}i\ep)[ -1 ]\\
DiF\,  \cO(-a\si_k) &=&iV_a^*   \ox \cO(\de_{k2}i\ep) [ 1 ]\\
FDiF\, \cO(-a\si_k) &=&\cO(-\ep + a(2\de_{k2}\ep + \si_k))\\
I \,   \cO(-a\si_k) = DFDiF\, \cO(-a\si_k)&=&\cO(\ep -2 a\de_{k2}\ep - a\si_k)
\eeqa

%\noindent
%{\large {\bf Auxiliary relations}}

{\em {\bf First auxiliary relation}}\footnote{here $\cO_{\si_k}$ 
are actually the sheafs extended by zero $\si_{k*} \cO_B$}
\beqa
\label{first}
F \, \cO(\al)&=&\cO_{\si_1}(\al - c_1) \, [ -1 ]\\
\label{second}
F \, \cO_{\si_k}&=&\cO(-\de_{k2}\ep)
\eeqa
For (\ref{first}) (cf. also 
equ. (2.48) of [\ref{ACHY}]) where $\al \in H^2(B)$ note that 
$F \, \pi^* \cO_B(\al)
=Rp_{1*}(p_2^* \pi^* \cO_B(\al)\ox \cP)
=\pi^* \cO_B(\al)\ox Rp_{1*}\cP$ where the second factor is
$Rp_{1*}\cP=R^1p_{1*}\cP [-1]$; the latter restricts on its support 
$\si_1(B)$ to $\si_1^* R^1 p_1 \cP = R^1 \pi_* \cP |_{\si_1\x_B Z}
=R^1\pi_* \cO_Z = (\pi_* \om_{Z/B})^*=\cO(-c_1)$, and 
%as $B$ is the support 
the assertion follows 
(so $F^1\cO = V_0$ as mentioned). For (\ref{second}) note 
$F \, \si_{k*}\cO_B=Rp_{1*}(p_2^*\si_{k*}\cO_B \ox \cP)$ and that
$p_2^*\si_{k*}\cO_B $ has support on $Z\x_B \si_k=Z$ (actually
the extension by zero of $\cO_Z=\pi^*\cO_B$ there) where 
(\ref{Poincare restrictions}) applies.

%\noindent
{\em {\bf Second auxiliary relation}}
\beqa
\label{plus a}
F \, \cO(a\si_k) &=&V_a^* \ox \cO(-\de_{k2}\ep - c_1)\\
\label{minus a}
F \, \cO(-a\si_k)&=&V_a   \ox \cO(-\de_{k2}\ep) [ -1 ]
\eeqa
For (\ref{minus a}) one gets for $a=1$ from the SES
\beqa
\label{start SES}
0 \lra \cO(-\si_k) \lra \cO \lra \cO_{\si_k} \lra 0
\eeqa
the following SES (coming from the LES of (\ref{start SES}) and 
using $F^0 \, \cO_Z =0$, $F^1 \, \cO_{\si_k} =0$) 
\beqa
0 \lra \cO(-\de_{k2}\ep) \lra F^1 \, \cO(-\si_k) \lra \cO_{\si_1}(-c_1) \lra 0
\eeqa
or, after tensoring with $\cO(-\si_1 + \de_{k2}\ep)$ 
and noting (\ref{adj rel}), (\ref{epsilon}),
\beqa
0 \lra \cO(-\si_1) \lra F^1 \, \cO(-\si_k) \ox \cO(-\si_1 + \de_{k2}\ep)
\lra \cO_{\si_1} \lra 0
\eeqa
identifying the middle term as $\cO$, i.e. giving indeed
$F^1 \, \cO(-\si_k)=V_1 \ox \cO(-\de_{k2}\ep)$. Similarly
one finds for the induction step that one gets from the SES
\beqa
0 \lra \cO(-(a+1)\si_k) \lra \cO(-a\si_k) \lra \cO_{\si_k}(ac_1) \lra 0
\eeqa
and the associated SES (coming from the LES with (\ref{second}))
\beqa
0 \lra \cO(ac_1 - \de_{k2}\ep) \lra F^1\, \cO(-(a+1)\si_k) 
\lra F^1\, \cO(-a\si_k) \lra 0
\eeqa
(using the induction hypothesis 
$F^1\, \cO(-a\si_k)= V_a \ox \cO(- \de_{k2}\ep) $) that
\beqa
0 \lra \cO(ac_1) \lra F^1\, \cO(-(a+1)\si_k) \ox \cO(\de_{k2}\ep) 
\lra V_a \lra 0
\eeqa
identifying indeed the middle term as $V_{a+1}$.
For (\ref{plus a}) one has in
an analogous induction from (\ref{start SES}) the 
SES $0\ra \cO \ra \cO(\si_k) \ra \cO_{\si_k}(-c_1) \ra 0$ and
from its LES
\beqa
0 \lra F^0 \cO(\si_k) \lra \cO(-\de_{k2}\ep - c_1) 
\lra \cO_{\si_1}(-c_1) \lra 0
\eeqa
(tensored with $\cO(\de_{k2}\ep + c_1)$) the $a=1$ case
$F^0 \cO(\si_k)= \cO(-\si_1 -\de_{k2}\ep - c_1)$. By the LES of the SES
$0 \ra \cO(a\si_k) \ra \cO( (a+1)\si_k) \ra \cO_{\si_k}(-c_1 - a c_1) \ra 0$
one concludes with (\ref{second}).

%\noindent
{\em {\bf Third auxiliary relation}}
\beqa
\label{rel 3}
F^1 \, \cO(\si_2-\si_1)=\cO_{D_2} \;\;\;\;\;\;\;\; (D_k:=2\de_{k2}\ep + \si_k)
\eeqa
with $D_2$ an effective divisor of the indicated class.
For this consider the SES
\beqa
0 \lra \cO(\si_2-\si_1) \lra \cO(\si_2) \lra \cO_{\si_{1}} \lra 0
\eeqa
which gives, from the LES, 
with $F^0 \, \cO(\si_2-\si_1)=0$,
$F^0 \, \cO(\si_2) = V_1^* \ox \cO(-\ep- c_1) 
= \cO(\si_2 - 2 \si_1 - 2 c_1)=\cO(-2\ep - \si_2)=\cO(-D_2)$, 
$F^0 \cO_{\si_1} = \cO$, $F^1 \, \cO(\si_2) =0$ the SES
\beqa
0 \lra \cO(-D_2) \lra \cO \lra 
F^1 \, \cO(\si_2-\si_1) \lra 0
\eeqa
 
{\em {\bf Fourth auxiliary relation}}
\beqa
\label{Xa}
X_a:=F^1 \, \Big( iV_a^*   \ox \cO(\de_{k2}i\ep)\Big)
=\cO\Big( -\ep + a(2\de_{k2}\ep + \si_k)\Big) 
\eeqa
The $a=1$ case follows 
with $iV_1=\cO(\si_2)$ and (\ref{minus a}) from
$X_1=F^1\Big( \cO(-\si_2+\de_{k2}(\si_2-\si_1+c_1)\Big)
=\cO(\si_k - \ep + 2 \de_{k2}\ep)$.
Then one gets from the $i$-transform of (\ref{Va}) the SES
\beqa
0 \lra iV_a^* \ox \cO(\de_{k2}i\ep) \lra iV_{a+1}^* \ox \cO(\de_{k2}i\ep) \lra 
\cO(\de_{k2}i\ep - a c_1) \lra 0
\eeqa
and then, from the LES, with (\ref{first}) and (\ref{rel 3}) 
for $k=1$ and $2$, the SES
\beqa
0 \lra X_a \lra X_{a+1} \lra \cO_{D_k}((-1)^k c_1-ac_1) \lra 0
\eeqa
After tensoring with $X_a^{-1}$, one gets a SES from
the inductive hypothesis
\beqa
0 \lra \cO \lra X_{a+1} \ox X_a^{-1}\lra \cO_{D_k}(D_k) \lra 0
\eeqa
identifying the middle term as $\cO(D_k)$ and 
concluding the induction (here we used 
$\cO_{D_k}\Big( (-1)^kc_1 - a c_1 + \ep - a ( \si_k + 2 \de_{k2}\ep)\Big)
=\cO_{D_k}\Big (D_k - a (c_1 + D_k)\Big)$
from $\ep = D_2 - \si_1 - c_1$ and $\si_1|_{D_2}=0$ for $k=2$).
Note here that $(D_2+c_1)D_2 = 6c_1 \ep$ vanishes on $C=n\si_1+\eta$.\\
{\em Remark:} Using restriction to $C$ we prove (\ref{I operation})
actually for $L=l|_C$ what is enough.

\section{Appendix: Relation between $A$- and $B$-model}

{\em Transitions}

We mention the following relation between the global $A$- and $B$-models.
First one has, as described above, 
the generic $A$-model with fibre ${\bf P^2}={\bf P_{111}}$
\beqa
A_{111}^{gen}\;\;\;\;\;\;\;\;\;\;\;\;\;\;\;\;\;\;\;\;\;\;\;
\;\;\;\;\;\;\;\;\;\;\;\;\;\;\;\;\;\;\;\;\;
zy^2 + x^3 + \alpha_4 xz^2 + \beta_6 z^3 =0
\eeqa
This has a specialisation $\alpha_4=b_4-a_2^2, \beta_6=-a_2b_4$
so that one gets
(with $w:=x-az$ of well-defined ${\cal L}$-weight)
\beqa
\label{A111spec}
A_{111}^{spec}\;\;\;\;\;\;\;\;\;\;\;\;\;\;\;\;\;\;\;\;\;\;\;\;\;\;\;\;\;\;\;\;
zy^2 + x^3 + (b_4 - a_2^2)z^2x - a_2b_4z^3 &=& \nonumber\\
zy^2 + (x-a_2z)(x^2+a_2xz+b_4z^2) &=& \\
zy^2 + w \Big(w^2 + 3a_2zw + (2a_2^2+b_4)z^2\Big) &=& 0\nonumber
\eeqa
By this specialisation a second section $\si_2$ is enforced at
$(w,y,z)=(0,0,1)$. From $\Delta = (4b_4 - a_2^2)(b_4 + 2 a_2^2)^2$ one
sees using Kodairas classifications of singular fibers that it acquires an 
$A_1$ fiber over a divisor $D$ of class $4 c_1$, more precisely
at $(w,y,z)=(0,0,1)$ over $2a_2^2+b_4=0$ (cf. last line of (\ref{A111spec})). 
If one resolves (by one blow-up)
the singularities over $D$ one gets an extremal transition from $Z$ 
to a model $\hat Z$ (considered in type IIA in [\ref{KM},\ref{KPM}]), 
with the Euler number changed [\ref{KLRY}] by 
$\delta=-2 e_D=24  c_1^2$, i.e. $e_Z=-36 c_1^2$. 
This model differs from the $B$-model over the same base by a 
birational transformation. 

To shed more light on the relation of the $A$- and $B$-model we point
to the following coordinate transformations. 
The affine form $y^2 + x^3 + \alpha_4 x + \beta_6  =0$
of $A_{111}^{gen}$ can be completed also in a ${\bf P_{231}}(6)$
\beqa
\label{A231gen}
A_{231}^{gen}\;\;\;\;\;\;\;\;\;\;\;\;\;\;\;\;\;\;
\;\;\;\;\;\;\;\;\;\;\;\;\;\;\;\;\;\;\;\;\;
y^2 + x^3 + \A_4xz^4 + \B_6z^6=0
\eeqa
Besides the former specialisation $\A_4 = b_4-a_2^2$ and $\B_6 = -a_2b_4$
which gives here 
\beqa
\label{A231spec}
A_{231}^{spec}\;\;\;\;\;\;\;\;\;\;\;\;\;\;\;\;\;\;\;\;\;\;\;\;\;\;
y^2 + x^3 + (b_4-a_2^2)xz^4 - a_2b_4z^6=0
\eeqa
one can also consider the specialisation $\A_4 = h_4+3f_2^2$ and 
$\B_6 = - (\frac{1}{4}g_3^2 + \frac{1}{27}f_2^3)$, i.e.
\beqa
\;\;\;\;\;\;\;\; A_{231}^{spec'}\;\;\;\;\;\;\;\;\;\;\;\;\;\;\;\;\;
y^2 + x^3 + (h_4+3f_2^2)xz^4 - (\frac{1}{4}g_3^2 + \frac{1}{27}f_2^3)z^6=0
\eeqa
which arises also after doing
$y\ra y - \frac{1}{2}g_3z^3, x \ra x - \frac{1}{3}f_2z^2$ in
\beqa
y^2 + x^3 + h_4 xz^4 + g_3 yz^3 + f_2 x^2z^2 = 0
\eeqa
The blow-up 
$x_{(2)} \ra u_{(2)}, y_{(3)} \ra u_{(2)}y_{(1)}$
(with projective weights indicated)
of the point $(x_{(2)}, y_{(3)})=(0,0)$ in the affine chart $z\neq 0$, 
with $y_{(1)}$ the new slope-coordinate along the
exceptional ${\bf P^1}$ given (locally and then also globally in the plane)
by $u=0$, which replaces the embedding plane 
${\bf P_{231}}$ by the new one ${\bf P_{121}} \ni (y,u,z)$, 
gives the global curve 
\beqa
u^2 + (f_2 z^2 + y^2 ) u + g_3 yz^3 + h_4z^4 =0
\eeqa
resp., after redefining $u \ra u - (fz^2+y^2)/2$,
the generic $B$-model of (\ref{Bmodel})
\beqa
B_{121}^{gen}\;\;\;\;\;\;\;\;\;\;\;\;\;\;\;\;
u^2 - \frac{1}{4}y^4 - \frac{1}{2}f_2 y^2z^2 
+ g_3yz^3 + (h_4 - \frac{1}{4}f_2^2)z^4 = 0
\eeqa

Finally we relate the specialisations $A_{111}^{spec}$ 
(or equivalently $A_{231}^{spec}$) and 
$A_{231}^{spec'}$. For this note that the respective blow-up's 
take place for $A_{111}^{spec}$ and still so for $A_{231}^{spec}$ at 
$(x,y,z)= (a,0,1)$ {\em over the curve} $2a_2^2+b_4=0$ in $B$ resp. for 
$A_{231}^{spec'}$ at $(x,y,z)= (0,0,1)$ even {\em in general}. 
Therefore a direct relation arises if
$2a_2^2+b$ vanishes not just on a curve in $B$ but generically; using
this further specialisation $b=-2a_2^2$ one gets from $A_{231}^{spec}$ 
\beqa
A_{231}^{further\, spec}\;\;\;\;\;\;\;\;\;\;\;\;\;\;\;\;
\;\;\;\;\;\;\;\;\;\;\;\;\;\;\;\;\;\;
y^2 + x^3 - 3 a_2^2 xz^4 + 2 a_2^3 z^6 =0
\eeqa
i.e. essentially just the form of $A_{231}^{spec'}$ when that is also
further specialised to $h=g=0$ 
(such that just the identification between $a_2$ and $f_2$ remains; 
this corresponds to the doubly specialised $B$-model
(cf. [\ref{ACK}]) with $b=0$ and $c=d^2$ in (\ref{Bmodel})). 

Alternatively one might also transform (the affine form of)
\beqa
B_{121}^{gen}\;\;\;\;\;\;\;\;\;\;\;\;\;\;\;\;
y^2 + x^4 + A_2 x^2 z^2 + B_3 x z^3 + C_4 z^4 =0
\eeqa
to the (affine) form of $A_{231}^{gen}$ in (\ref{A231gen})
with 
\beqa
\A_4 = -\frac{1}{4} \frac{1}{12} (A^2 + 12 C)\;\;\; , \;\;\;
\B_6 = -\frac{1}{4} 
\Big( \frac{1}{216} A (36 C - A^2) - \frac{1}{16}B^2 \Big)
\eeqa
For the special case of $B_3=0$ one gets the special form
$A_{231}^{spec}$ of (\ref{A231spec}) with
\beqa
a_2 = - \frac{1}{6}A_2 \;\;\; ,  \;\;\; 
b_4 = \frac{1}{144}(A_2^2 - 36 C_4)
\eeqa

{\em Specific divisors}

As mentioned it is important to understand the geometry of
the classes $D_k= R_{\ep} ( \si_k)=\si_k + 2\de_{2k}\ep$, e.g.
$D_2= \si_2 + 2 \ep$ did not come from a smooth {\em irreducible}
effective divisor.

Actually, in the (resolved specialised) 
$A$-model (cf. also [\ref{DOPW1999}]) 
$\si_2$ and $2\ep$ can be represented
individually by smooth irreducible surfaces for
$2\ep$ is represented by a surface
$\E$, rationally ruled by the fibre $\th$ over the curve 
$M=\si_2 \cdot \E$ in $\si_2$ of class $4c_1\si_2$ 
\beqa 
 & \E & \nonumber\\ 
D_2=\si_2 + \E \;\;\;\;\;\;\;\;\;\;\;\mbox{where} 
\;\;\;\;\;\;\;\;\;\;\;\;\;\;\;\;\; 
\pi_{\E}  & \da   & 
\;\;\;\;\; \mbox{fibre} \; \th \\ 
\;\;\;\;\;\;\;\;\;\;\;\;  \si_2 \cdot \E = & M & \nonumber 
\eeqa
Concerning the fibration of\footnote{the class $x\si_1 + y \si_2 + z c_1$
of the blow-up surface $\E$ is identified as $2\ep$ from the relations 
$\E \, F =0$, 
$\E \, \al \si_1 =0$ (this constrains also already 
the base curve in the ansatz)
and $\E \, \th = -2$ 
(the last relation from the $A_1$ resolution; 
for an a posteriori check cf. below)}
$\E$ over $M$ by $\th$
note that the surfaces $\E$ and $4c_1$ cut out in $\si_2$
the same curve $M$; if $\E = M \ " \x " \th$ the intersection
of $4c_1$ and $\E$ itself is $M^2|_{\si_2}=16c_1^2$ 
times the fibre but this is now indeed 
$\th$ because $\th =\frac{1}{2c_1^2}c_1\ep = \frac{1}{16c_1^2}4c_1\cdot \E$;
furthermore $\th^2|_{\E} =0$ ($\th$ is a fibre in $\E$) as 
$\th = \frac{1}{4c_1^2}c_1 \cdot \E$ gives 
$\th^2|_{\E} = \frac{1}{(4c_1^2)^2}c_1^2 \cdot \E 
= \frac{1}{16c_1^2}F \cdot \E = 0$;
finally $\th \cdot \E=\frac{1}{2c_1^2}c_1 2\ep^2=
\frac{1}{c_1^2}c_1 (-2c_1\si_2+c_1\ep)=-2$ so $\th$ is just a ${\bf P^1}$ 
(cf. the analogous case
of $(-2)$-curves in a $K3$; here $TZ|_{\th}=T\th \o+ N_{\E} \th \o+ N\E|_\th$
where the middle term is trivial because of the fibration).
Note further that the self-intersection number $-e:=M^2|_{\E}$ of the base 
of the fibration is $\si_2^2\E=-4c_1^2$. The class 
$c_1(\E)=2M+(e_M+e)\th$ (cf. 
[\ref{Hartsh}]) is then identified as $c_1(\E)=2M+(-12 c_1^2 + 4 c_1^2)\th
=2M - 8 c_1^2 \th$ verifying the adjunction relation 
$\E(\E + c_1(\E))=0$ for the irreducible surface $\E$ in the Calabi-Yau $Z$
as $\E|_{\E}=(-2\si_2 + 2 c_1)|_{\E}= -2M + 8 c_1^2 \th$.

There is a corresponding surface $\iota \E$
rationally ruled by $F-\th = \iota \th$ (as $\iota \ep = - \ep + 2 c_1$)
over the base curve $\iota M$ in $\si_1$. 
One has 
$\E \cdot \iota \E = 4 \ep ( - \ep + 2 c_1 )
= 4c_1 (\si_1 + \si_2 + c_1 )
=4c_1 (2\si_2 + \ep )
=8(c_1\si_2 + c_1^2 \th)$, 
the proper transform of the curve of singularities in the $A$-model.
Fibrewise the affine $A_1$-tree (two ${\bf P^1}$ intersecting twice)
is built by $\th$ and $\iota \th$ as 
$\th \cdot \iota \E = \frac{1}{16c_1^2}4c_1 \E \cdot \iota \E = 
F (\si_1 + \si_2 + c_1 ) =2$.

Note that in the $B$-model not only 
the term\footnote{whose non-vanishing implies that $D_2$ 
(being numerically effective) is not 
{\em irreducible} effective (it would be then a section)} 
$(D_2+c_1)D_2=6c_1\ep$
vanishes on surfaces of class $n\si_1+\al$ but that 
even $D_2$ itself is, on such a surface, 
given (as a class) just by its part on the 
elliptic surface\footnote{which is, for example, just $b \x F$ resp. $K3$ 
for $\al = b$ resp. $f$ 
($\al$ denotes both $\al \in H^2(B)$ and $\pi^* \al$)} 
$\al$
as $D_2|_{n\si_1 + \al}= D_2|_{\al}$. 
What concerns the self-intersection number of $D_2|_{\al}$ in $\al$
note that only the $\si_2$ part of $D_2$ contributes as
$(D_2|_{\al})^2=D_2^2\al=4c_1\al\ep - c_1\al \si_2=- c_1\al \si_2
=\al \si_2^2 = (\si_2|_{\al})^2$.
Inside ${\al}$ now $D_2$ can be\footnote{Contrast this also 
with the fact (cf. for example [\ref{hor and vert}])
that, inside the elliptic $K3$ given by $\pi^*f$,
the numerical section $D:=\si_2 + p F$ can not be represented by
an irreducible curve or else it would be a section, so rational
but one has $D^2|_f= - c_1\si_2 |_f + 2p = -2 + 2p \neq -2$ for $p\neq 0$.}
an irreducible section: 
one has $(D_2+c_1)D_2|_{\al}=0$ and
its $K_{D_2|_{\al}}= K_{\al}\cdot D_2|_{\al} + (D_2|_{\al})^2
=\al \cdot D_2 \cdot \al - c_1 \al \si_2 = (\al - c_1)\al \si_2$
equals the the $K$ of the base curve as 
$K_{\al} = K_B \, \al + \al^2 = \al^2 - c_1 \al$ (computed in $B$; here $\al$
refers to the class in $H^2(B)$ whereas usually it denotes $\pi^* \al$).

\section{Restrictions on the bundle parameters}

Here we describe conditions on the bundles which appear besides
the generation-tuning and the involution-invariance like
effectivity of the five-brane class occurring
in the anomaly cancellation condition.

\subsection{\label{uppbou}The upper bound on $\eta$}

We restrict our attention 
to the visible sector concerning $V_1$ (which we simply call $V$) 
and put in the hidden sector $V_2=0$ 
when embedding the bundle $(V_1,V_2)$ in $E_8\times E_8$.
Essential restrictions on $V$ come from bounds on the $\eta$ class. 
The upper bound comes from the anomaly cancellation condition
$c_2(Z)=c_2(V_1)+c_2(V_2)+W$ giving
the effectiveness restriction $c_2(V)\le c_2(Z)$ on the
five-brane class $W=W_B+a_f \, F$ 
(as the five-brane must wrap an actual curve).
We want to make sure that $W_B$ is effective and $a_f$ is non-negative.

For the $B$-model the decomposition $(\oplus_i \sigma_i H^2(B))\oplus H^4(B)$
(with suitable pull-backs understood)
gives (with $\eta_2=0$) for the parts not coming from $H^4(B)$ 
\beqa
\label{Weff}
\eta\sigma_1-\mu \eta (\si_1-\si_2)+W_B=6c_1\sigma_1+6c_1\sigma_2
\eeqa
{\em The $\mu=0$ sector}

In the final result for 
$c_2(V)=\eta\sigma_1+\omega$ 
(where $\omega\in H^4(B)$, pulled back to $Z$) the 'number' $k$ 
of the model ($1,2$ for $A,B$) cancelled out 
leaving the $A$ model result unchanged (cf. previous footn.). 
On the other hand note that we have a corresponding decomposition 
$c_2(Z)=\frac{12}{k}\Sigma c_1 + (c_2+(\frac{12}{k}-1)c_1^2)$).
This is the term responsible
for the upper bound $\eta\sigma_1\le \frac{12}{k}\Sigma c_1$, which
when interpreted as the sufficient condition
$\eta\le \frac{12}{k}c_1$ is thus for $k=2$ in the $B$-model
much sharper than for $k=1$ in the $A$-model, excluding for example
$(x,y)=(14,22)$ and $(24,30)$ over ${\bf F_2}$.

But actually the situation is slightly better because
we established a sufficient condition only. The base part (\ref{Weff})
of the condition gives for $\mu=0$ certainly the sufficient condition 
$\eta_1 = xb+ yf \le 6c_1 = 12 b + 24 f $. But because the curve classes
are not all independent there is actually more flexibility. 
For this let us restrict our attention 
to the phenomenologically relevant case 
of the base ${\bf F_2}$. 
Then $6c_1 \si_1 + 6c_1 \si_2=(24b+24f)\si_1 + 24f\si_2$ 
and the condition
$\eta=xb+yf \le 24b+24f$ ensues allowing $(x,y)=(14,22)$. 

{\em The $\mu\neq 0$ sector}

With $\mu = 1/2$ to get invariant bundles 
(cf. below) (\ref{Weff}) amounts to
\beqa
\frac{1}{2}\eta \si_1 +\frac{1}{2}\eta \si_2 \le 6c_1\sigma_1+6c_1\sigma_2
\eeqa
Even without exploiting the less restrictive version of the 
bound as above, one is back to the $A$-model bound $\eta \le 12 c_1$ 
as sufficient, allowing $(x,y)=(14,22)$ and $(24, 30)$.

\subsection{The lower bound on $\eta$}

This is a bound on 'how much instanton number has to be turned on
to generate/fill out a certain $SU(n)$ bundle, i.e.
to have no greater unbroken gauge group than a certain $G$'.

Recall that in the six-dimensionsional duality [\ref{MVII}] 
between the heterotic string on $K3$ with
instanton numbers $(12\mp m)$ (no five-branes) and $F$-theory on
$F_m$ the gauge group is 
described by the singularities of the $F$-theory fibration; a perturbative
heterotic gauge group corresponds to a degeneration over the common
base-${\bf P^1}$ $B_1=b$ of the heterotic $K3$ resp. the $F_m$, 
i.e. the discriminant divisor 
$\Delta=12c_1(F_m)$ has a component $\delta(G)\, b$ (with $\delta(G)$
the vanishing order of the discriminant, equivalently the Euler number of
the affine resolution tree of the singularity). This gives the relation
$m\le \frac{24}{12-\delta(G)}$ for having not a singularity worse than $G$ 
[\ref{R}]
(as $\Delta^{\prime}=\Delta-\delta(G)b$ has transversal intersection 
with $b$ and so $\Delta^{\prime}\cdot b\ge 0$), i.e. 
$12-m\ge 12-\frac{24}{12-\delta(G)}=(6-\frac{12}{12-\delta(G)})c_1(B_1)$.  
From 
this was induced [\ref{R}] (cf. [\ref{BM}])
the bound in four dimensions 
\beqa
\eta_1\ge (6-\frac{12}{12-\delta(G)})c_1
\eeqa
(the $(12\mp m)$ structure generalizes in four dimensions to 
$\eta_{1/2}=6c_1\mp t$ from
the component part
$\eta_1\sigma+\eta_2\sigma=12c_1\sigma$ 
of the anomaly cancellation condition $c_2(V_1)+c_2(V_2)+a_fF=
c_2(Z)$ for the case of an $A$-model with $W_B=0$). 
In our case 
($B$-model at the invariant point $\mu = 1/2$)
(\ref{Weff}), including an $\eta_2\neq 0 $, similarly
reads $(\eta_1+\eta_2)\Sigma = 12 c_1 \Sigma$ 
(for the case with $W_B =0$; here $\Sigma = \sigma_1+ \sigma_2$).

One gets\footnote{As our focus in this paper is 
on the invariance condition rather than on specific $\eta$-classes 
we just mention the 
similar criterion $\eta \geq 5c_1$ (and $\eta\, b \geq 0$) [\ref{OPP}].}
the lower bound $\eta \ge \frac{30}{7} c_1$ 
for an $SU(5)$ unbroken gauge group (with $\delta=5$).

\subsection{Examples}

So one has to find an $\eta$
in the strip 
$\frac{30}{7}c_1\le \eta \le 6c_1$ with $\eta(\eta-5c_1)=\pm 12$
(for $\lambda=\pm \frac{1}{2}$) resp. $\eta(\eta-5c_1)=\pm 4$ 
(for $\lambda=\pm \frac{3}{2}$) with $a_f$ non-negative (we will be
considering only $F_0,\, F_1,\, F_2$) where
$a_f=c_2+10c_1^2$ for $\lambda=\pm \frac{1}{2}$ resp.
$a_f=c_2+10c_1^2\mp 20$ for $\lambda=\pm \frac{3}{2}$.

$6$ generations bundles over\footnote{cf. also [\ref{DOPW1999}]; 
data which specify a $6$ generations bundle over $F_1$ 
(where no involution was found) 
and fulfill the bounds mentioned above 
are given by
$\eta= 11b+15f\; , \;\lambda=+3/2$.
A $6$ generations bundle over ${\bf F_0}$ with $\eta = 8b + 14 f,  
\lambda = + 3/2$ violates the lower bound $x,y \ge 9$ (resp. $10$).} 
$F_2$ are
(violating the (naive) upper bound $x \le 12, y \le 24$)
\beqa
\eta =14b+22f \;\;\; , \;\;\; \lambda = + 3/2 \;\;\;\;\;\;\;\;\;\; 
; \;\;\;\;\;\;\;\;\;\;
\eta =24b+30f \;\;\; , \;\;\; \lambda = - 1/2
\eeqa 
with
$W=(5b+13f)(\si_1+\si_2) + 75 F$ and $W=9f(\si_1+\si_2) + 99 F$, 
respectively (from $c_2(V_{\mu=1/2})=\frac{1}{2}\eta(\si_1+\si_2) 
-\frac{1}{4}\eta c_1 -40 + \mbox{sign}(\lambda)(|\lambda|-1/2)20$ 
%with the last term appearing
%only in the case $\lambda = \pm 3/2$ but being absent for $\lambda = \pm 1/2$)
.
Note that in both cases $W_B$ is not only effective 
(the coefficients of 
$b$ and $f$ are nonnegative) but the class will be represented also by a 
(smooth) {\em irreducible} curve (so that one will not have to worry about
intersections of its components) as $W_B \cdot b \geq 0$ 
(cf. [\ref{Hartsh}]). (But note also $\eta b < 0$.)

\section*{References}
\begin{enumerate}

\item
\label{W}
E. Witten, {\em New issues in manifolds of $SU(3)$ holonomy}, 
Nucl. Phys. {\bf B268} (1986) 79.

\item
\label{FMW}
R. Friedman, J. Morgan and E. Witten, {\em Vector bundles and $F$-theory},
Comm. Math. Phys. {\bf 187} (1997) 679, hep-th/9701162.

\item
\label{C}
G. Curio, {\em Chiral matter and transitions in heterotic string models},
Phys. Lett. {\bf B435} (1998) 39, hep-th/9803224.

%\item
%\label{Wsy}
%E.Witten, {\em Symmetry breaking patterns in superstring models}, 
%Nucl. Phys. {\bf B258} (1985) 75.

\item
\label{ACK}
B. Andreas, G. Curio and A. Klemm,
{\em Towards the Standard Model from elliptic Calabi-Yau},
Int.J.Mod.Phys. {\bf A19} (2004) 1987, hep-th/9903052.

%\item
%\label{KLM}
%A.Klemm, W.Lerche and P.Mayr, 
%{\em K3--Fibrations and Heterotic-Type II String Duality},
%Phys. Lett. {\bf B357} (1995) 313, hep-th/9506112.

\item
\label{KLRY}
A. Klemm, B. Lian, S.-S. Roan and S.-T. Yau, 
{\em Calabi-Yau fourfolds for M- and F-Theory compactification}, 
Nucl. Phys. {\bf B518} (1998) 515, hep-th/9701023.

\item
\label{BKMT}
P. Berglund, A. Klemm, P. Mayr and S. Theisen, 
{\em On Type IIB Vacua With Varying Coupling Constant}, 
Nucl.Phys. {\bf B558} (1999) 178,
hep-th/9805189.

\item
\label{KM}
A.\ Klemm and P.\ Mayr, {\em Strong Coupling Singularities and 
Non-abelian Gauge Symmetries in $N=2$ String Theory},
Nucl.Phys. {\bf B469} (1996) 37-50.

\item
\label{KPM}
S.\ Katz, D.\ Morrison and R.\ Plesser  {\em Enhanced Gauge Symmetry in 
Type II String Theory}, Nucl.Phys. {\bf B477} (1996) 105-140.

\item
\label{MVII}
D.\ Morrison and C.\ Vafa, {\em 
Compactifications of F-Theory on Calabi--Yau Threefolds -- II},
Nucl.Phys. {\bf B476} (1996) 437-469. 

\item
\label{DLOW}
R. Donagi, A. Lukas, B.A. Ovrut and D. Waldram, {\em Holomorphic 
vector bundles and non-perturbative vacua in M-theory},
JHEP {\bf 9906} (1999) 034, hep-th/9901009.

\item
\label{CD}
G. Curio and R. Donagi, Nucl. Phys. {\bf B518} (1998) 603, hep-th/9801057.

\item
\label{BM}
P. Berglund and P. Mayr, {\em Heterotic String/F-theory Duality 
from Mirror Symmetry}, Adv.Theor.Math.Phys. {\bf 2} (1999) 1307,
hep-th/9811217.

\item
\label{R}
G. Rajesh, {\em Toric Geometry and $F$-theory/heterotic duality in four
dimensions}, hep-th/9811240.

\item
\label{DOPW1999}
R. Donagi, B.A. Ovrut, T. Pantev and D. Waldram,
{\em Standard Models from Heterotic M-theory},
Adv.Theor.Math.Phys. {\bf 5} (2002) 93, hep-th/9912208.

\item
\label{DOPW2000}
R. Donagi, B.A. Ovrut, T. Pantev and D. Waldram,
{\em Standard-Model Bundles on Non-Simply Connected Calabi--Yau Threefolds},
JHEP {\bf 0108} (2001) 053, hep-th/0008008;
{\em Standard-model bundles},
Adv.Theor.Math.Phys. {\bf 5} (2002) 563, math.AG/0008010;
{\em Spectral involutions on rational elliptic surfaces},
Adv.Theor.Math.Phys. {\bf 5} (2002) 499, math.AG/0008011.

\item
\label{OPP}
B.A. Ovrut, T. Pantev and J. Park,
{\em Small Instanton Transitions in Heterotic M-Theory},
JHEP {\bf 0005} (2000) 045, hep-th/0001133.

\item
\label{ACHY}
B. Andreas, G. Curio, D. Hernandez Ruiperez, S.-T. Yau
{\em Fourier-Mukai Transform and Mirror Symmetry 
for D-Branes on Elliptic Calabi-Yau},
math.AG/0012196;
{\em Fibrewise T-Duality for D-Branes on Elliptic Calabi-Yau},
hep-th/0101129, JHEP {\bf 0103} (2001) 020.

\item
\label{Hartsh}
R. Hartshorne
{\em Algebraic Geometry},
Graduate Texts in Mathematics (1977) Springer-Verlag.

\item
\label{hor and vert}
B. Andreas and G. Curio,
{\em Horizontal and Vertical Five-Branes in Heterotic/F-Theory Duality},
hep-th/9912025, JHEP {\bf 0001} (2000) 013.

\item
\label{Andreas}
B. Andreas,
{\em On Vector Bundles and Chiral Matter 
in N=1 Heterotic Compactifications},
hep-th/9802202, JHEP {\bf 9901} (1999) 011.

\end{enumerate}
\end{document}